# Optical and Microcavity Modes Entanglement by means of Developed Opto-Mechanical System

Ahmad Salmanogli[1,2], H. Selçuk Geçim[1]
[1]Faculty of Engineering, Electrical and Electronics Engineering Department, Çankaya University, Ankara, Turkey
[2] Chemical Engineering Department, Bioengineering faculty, Hacettepe University, 06800, Ankara, Turkey.

*Abstract*— To improve the performance of the traditional tripartite opto-mechanical system, we design a developed opto-mechanical tripartite system in which all subsystems such as an optical cavity, microresonator, and microcavity modes are coupled directly to each other. This system is analyzed quantum mechanically, and its dynamics of motion is studied by the Heisenberg-Langevin equations. Notably, the originality of this work contributes to embedding the spherical capacitor in the optical cavity where the plasmon-plasmon interaction between capacitor's plates manipulates the internal electric field. Therefore, using plasmon-plasmon interaction inside the spherical capacitor, which leads to nonlinearity, causes the contributed polarization function of the plasmonic property. Actually, this plasmonic property manages the spherical capacitance. Although the spherical capacitor is fundamentally dependent on the plasmonic properties caused by the capacitor's gap, the spherical capacitor for a design is required to be parallel to the microwave capacitor. The system developed in the present work proved that the stability of the entanglement between optical cavity and microcavity modes toward the temperature changing is dramatically increased; which is the original goal of this study. It is worth mentioning that the environment noise severely restricted the traditional tripartite system operated with a wide bandwidth, hence the present design can efficiently fix this issue.

## I. INTRODUCTION

Recently, fabrication of sensitive sensors has been a subject of intense research, which may lead to replacing the classical sensors by the quantum ones. For example, one can consider the classical radar system replaced by the quantum radar system [1, 2]. Quantum radar generally operates with the concept of radar, even though this device operation is contributed to a relatively small number of photons. The motivation for quantum radar developing is that using quantum phenomena allows increasing the sensitivity of the sensing function. Nonetheless, quantum radar uses explicitly and implicitly some of the advantages of classical radar systems. For instance, microwaves photons penetration is very good in clouds, while visible photons are most likely to be absorbed or scattered in the related medium [1].

One important element that is normally used in quantum radar is the opto-mechanical system (tripartite system), which can generate the entangled photons. In this system, the priority is the generation of the entangled photons [3-8] between optical cavity (OC) photons and microwave (MW) cavity photons. The tripartite system, which contains OC, MW cavity, and microresonator (MR), is a useful system that can be used to generate the entangled outputs between different modes [9]-[12]. This system has the ability by which the MW modes are entangled with the OC modes, due to this fact, the system sensitivity performance will be dramatically enhanced [1]. For instance, the enhancement of the photodetection sensitivity via quantum illumination was investigated in [2]. Moreover, photodetection of the propagating quantum microwaves in circuit QED is studied. Additionally, several interesting works have been reported about the coupling MR mode with optical cavity OC mode and MW mode [9], [10]. In these works, MR can efficiently be coupled with MW mode through capacitor or inductor and can be coupled with QC either optical pressure or optical gradient force. For this reason, generally, a tripartite system is defined where the quantum interface between subsystems is transferred with a high fidelity, for which some works have focused on the entanglement between the subsystems modes. In the traditional tripartite system, it has been usually shown that MR can be simultaneously coupled with OC and MW modes. In this system, commonly, the vibrating drumhead is capacitively coupled with the MW cavity modes and coupled with the optical cavity through optical pressure. In the recent works, authors have mainly tried to show how the entanglement between the MR-OC, MR-MW, and MW-OC can be properly defined. For these reasons, many different ideas have been proposed to found that (I) the system the quantum interface has ability to transfer quantum state with high fidelity [9], (II) the efficacy the MR can be coupled with either MW or OC [9], (III) the parameters that manipulate the system stability [10], [11], and (IV) the entanglement between movable mirrors and cavity fields [12]. However, in most published works, the operation temperature is very low because by increasing the environment temperature the noise applied to the MR system is very high due to its low operation frequency. Therefore, the entanglement between different modes is severely affected. In other words, increasing noise in MR system is coupled capacitively with MW cavity modes and the optical pressure to OC modes. The mentioned problem is addressed in some works by engineering the MR resonance bandwidth [11]. The authors of the present study are of the opinion that the nanoelectronic-based method can be used because of its good degree freedom for system manipulation. Accordingly, we developed a new OC in which a spherical capacitor is embedded into OC. The spherical capacitor with two noble metal plates can generate the internal plasmonic field to introduce the nonlinearity into the medium. The generated plasmonic field contributes to the plasmon-plasmon interaction [13, 14], [17, 18]. This field in the gap region between plates is so high that it can cause a considerable nonlinearity [15], [16]. In fact, by this system, we want to directly couple the OC modes with MW cavity modes while in the traditional system the coupling between MW and OC was indirect through the MR modes. Moreover, the embedded capacitor in the equivalent electronic system is parallel with the MW circuit capacitor by which we can control the temperature influence on the output entanglement. In the following, the theoretical background of this system, which is analyzed quantum mechanically, is presented.

## II. THEOREY AND BACKGROUNDS

### A. System Illustration

In this section, we introduced the analysis of the developed tripartite system, which is schematically illustrated in Fig. 1. Fig. 1a presents the traditional opto-mechanical tripartite system containing MW, OC, and MR. It is schematically shown that OC mode can be coupled with MR through optical pressure and MR can capacitively be coupled with MW cavity mode. The developed version of the tripartite system is illustrated in Fig. 1b. We embedded a spherical capacitor in OC. Here, using the plasmon-plasmon interaction between metal plates induces a nonlinear effect. Therefore, the spherical capacitance in OC medium is severely affected by the capacitor gap region nonlinearity and plasmon-plasmon interaction. The equivalent electronic circuit for the traditional tripartite system and developed one is depicted in Fig. 1c. As can be seen, the spherical capacitor is parallel with MW circuit capacitor which means that the total capacitance in the developed system is $C(x) + C(P)$, where $C(x)$ is an alterable capacitor dependent on MR system and $C(P)$ is related to the second mode harmonic generated by the nonlinearity in spherical capacitor gap region. It is noteworthy that MW cavity mode is coupled directly with OC modes. In fact, our work originality contributes to system operation at high temperatures since the MR is severely affected by that. Although the $C(x)$ is dramatically influenced by the input noise, the total capacitance is not affected where $C(P)$ can be a dominant factor but it is slightly influenced by the environment temperature. Hence, in the developed system, the alteration of $C(x)$ can be considered as a small signal around the operation point, which is determined by the $C(P)$.

### B. System Dynamics

Based on the developed system, we assumed that the OC is simultaneously coupled to MR through optical pressure and to the MW via spherical capacitance. In addition, MR is coupled to MW through the $C(x)$. Therefore, the Hamiltonian of the developed tripartite system is given by:

$$H_{total} = \frac{p_x^2}{2m} + \frac{m\omega_m^2 x^2}{2} + \frac{\varphi^2}{2L} + \frac{Q_x^2}{2C(x)} + \frac{Q_p^2}{2C(P)} - e(t)Q_x + \hbar\omega_c a_1^+ a_1 - \hbar G_{oc} a_1^+ a_1 x$$
$$+ \hbar\chi^{(2)} E_f (a_2^+ a_1 a_1 + a_1^+ a_1^+ a_2) + i\hbar E_c (a_1^+ e^{(-j\omega_{oc} t)} - a_1 e^{(j\omega_{oc} t)}) \quad (1)$$

where ($p_x$, x) are the canonical position and momentum of MR with frequency $\omega_m$, ($\varphi$, $Q_x$) are the canonical coordinates for MW cavity, ($\varphi$, $Q_p$) are the canonical coordinates for the second harmonics generated in OC cavity, and $\varphi$, L, $Q_x$, and $Q_p$ are the inductor flux, inductance in Henrys, charge on the capacitor $C(x)$ and charge on the spherical capacitor $C(P)$, respectively. Moreover, $a_1$ and $a_1^+$ are the annihilation and creation operator for OC which is pumped with frequency $\omega_{oc}$, and ($a_2$, $a_2^+$) are the second harmonics generation annihilation and creation modes with the second order susceptibility ($\chi^{(2)}$), respectively. Finally, $E_c$, $e(t)$, $G_{o\omega}$, and $G_{oc}$ are the input driving laser, MW cavity driving, coupling factor between MW and MR, and coupling factor between OC and MR [9], respectively. However, it should be noted that in the system Hamiltonian, for the sake of simplicity, the interaction between cavity mode and charges on the spherical capacitor is ignored. In the following, using some dimensionless position and momentum operator of MR and also by defining raising and lowering operator for MW cavity [9]-[12], Eq. 1 can be re-written as:

$$H_{total} = \frac{\hbar\omega_m(p^2+q^2)}{2} + \hbar\omega_\omega b^+b - \hbar G_{o\omega}b^+bq + \hbar\omega_c a_1^+a_1 - \hbar G_{oc}a_1^+a_1q + \hbar\omega_{c2}a_2^+a_2 - \hbar G_{oc2}a_2^+a_2q$$
$$+ \hbar\chi^{(2)}E_f(a_2^+a_1a_1 + a_1^+a_1^+a_2) + i\hbar E_c(a_1^+e^{(-j\omega_{oc}t)} - a_1e^{(j\omega_{oc}t)}) + +i\hbar E_\omega(b^+e^{(-j\omega_{o\omega}t)} - be^{(j\omega_{o\omega}t)}) \quad (2)$$

where $E_\omega$ relates to the MW cavity driving, ($b^+$, $b$) are the MW cavity annihilation and creation operator, and (q, p) is the dimensionless position and momentum operator of MR subsystem, respectively [9]. For simplicity, one can ignore the fast oscillating terms such as $\pm\omega_{oc}$ and $\pm\omega_{o\omega}$ so that the final form of the system Hamiltonian becomes:

$$H_{total} = \frac{\hbar\omega_m(p^2+q^2)}{2} + \hbar\Delta_\omega b^+b - \hbar G_{o\omega}b^+bq + \hbar\Delta_c a_1^+a_1 - \hbar G_{oc}a_1^+a_1q + \hbar\Delta_{c2}a_2^+a_2 - \hbar G_{oc2}a_2^+a_2q$$
$$+ \hbar\chi^{(2)}E_f(a_2^+a_1a_1 + a_1^+a_1^+a_2) + i\hbar E_c(a_1^+ - a_1) + +i\hbar E_\omega(b^+ - b) \quad (3)$$

where $\Delta_\omega = \omega_\omega - \omega_{0\omega}$, $\Delta_c = \omega_c - \omega_{0c}$, and $\Delta_{c2} = \omega_{c2} - 2\omega_{0c}$. However, the dynamics of the contributed modes in this system are also affected by the damping and noise operator since each mode interacts with its own environment. In the following, the system equations of motions by using Heisenberg-Langevin equation are presented as:

$$\dot{q} = \omega_m p$$
$$\dot{p} = -\omega_m q - \gamma_m p + G_{o\omega}b^+b + G_{oc}a_1^+a_1 + G_{oc2}a_2^+a_2 + \xi \quad (4)$$
$$\dot{a}_1 = -(i\Delta_c + \kappa_c)a_1 + iG_{oc}qa_1 - i\chi^{(2)}E_f a_2^+(2a_1^+) + E_c + \sqrt{2\kappa_c}\varepsilon_c$$
$$\dot{a}_2 = -(i\Delta_{c2} + \kappa_{c2})a_2 + iG_{oc2}qa_2 - i\chi^{(2)}E_f a_1^2 + \sqrt{2\kappa_{c2}}\varepsilon_{c2}$$
$$\dot{b} = -(i\Delta_\omega + \kappa_\omega)b + iG_{o\omega}qb + E_\omega + \sqrt{2\kappa_\omega}\varepsilon_b$$

where $\gamma_m$, $\xi$, $\varepsilon_c$, $\varepsilon_{c2}$, and $\varepsilon_b$ are the mechanical damping rate, quantum Brownian noise acting on MR, OC cavity input noise, and MW cavity input noise, respectively. As can be seen, Eq. 4 is a nonlinear equation. To linearize this equation, one can use the fluctuations associated with the field modes such as $q = \langle q \rangle + \delta q$, where $\langle q \rangle$ stands for field average in the steady-state condition and $\delta q$ indicates the fluctuation of the considered mode [9]-[12]. Using this term, we can obtain the equations of motion based on the fluctuation in the form of linearization approximation as:

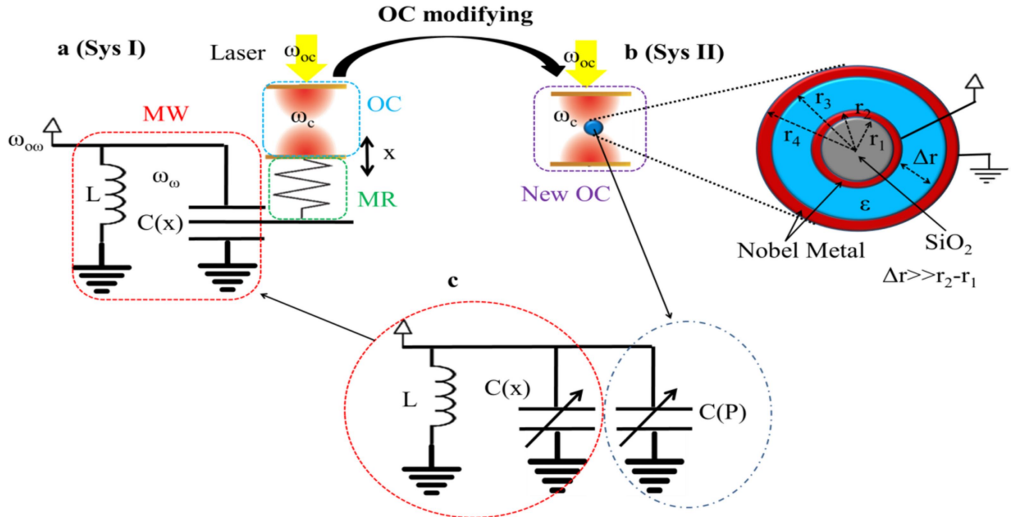

Fig.1. Developed opto-mechanical tripartite system schematic: (a) traditional system, (b) developing with inserting a spherical capacitor in optical cavity, (c) Electronics equivalent circuit.

$$\dot{\delta q} = \omega_m \delta p$$
$$\dot{\delta p} = -\omega_m \delta q - \gamma_m \delta p + G_{o\omega}[\beta^*\delta b + \beta\delta b^+] + G_{oc}[\alpha_1^*\delta a_1 + \alpha_1\delta a_1^+] + G_{oc2}[\alpha_2^*\delta a_2 + \alpha_2\delta a_2^+] + \xi \quad (5)$$
$$\dot{\delta a}_1 = -(i\Delta_c + \kappa_c)\delta a_1 + iG_{oc}[q_1\delta a_1 + \alpha_1\delta q] - 2i\chi^{(2)}E_f[\alpha_1^*\delta a_2 + \alpha_2\delta a_1^+] + \sqrt{2\kappa_c}\varepsilon_c$$
$$\dot{\delta a}_2 = -(i\Delta_{c2} + \kappa_{c2})\delta a_2 + iG_{oc2}[q_1\delta a_2 + \alpha_2\delta q] - 2i\chi^{(2)}E_f\alpha_1\delta a_1 + \sqrt{2\kappa_{c2}}\varepsilon_{c2}$$
$$\dot{\delta b} = -(i\Delta_\omega + \kappa_\omega)\delta b + iG_{o\omega}[q_1\delta b + \beta\delta q] + \sqrt{2\kappa_\omega}\varepsilon_b$$

where $q_1 = \langle q \rangle$, $p_1 = \langle p \rangle$, $\alpha_i = \langle a_i \rangle$ (i = 1, 2), and $\beta = \langle b \rangle$, which are the contributed modes average. These parameters are calculated through the steady state equations, which are introduced as:

$$p_1 = 0$$
$$-\omega_m q_1 - \gamma_m p + G_{o\omega}\beta^*\beta + G_{oc}\alpha_1^*\alpha_1 + G_{oc2}\alpha_2^*\alpha_2 = 0$$
$$-(i\Delta_c + \kappa_c)\alpha_1 + iG_{oc}q_1\alpha_1 - 2i\chi^{(2)}E_f\alpha_1^*\alpha_2 + E_c = 0 \quad (6)$$
$$-(i\Delta_{c2} + \kappa_{c2})\alpha_2 + iG_{oc2}q_1\alpha_2 - i\chi^{(2)}E_f\alpha_1^2 = 0$$
$$-(i\Delta_\omega + \kappa_\omega)\beta + iG_{o\omega}q_1\beta + E_\omega = 0$$

Since Eq. 5 is not linear, it needs to be solved numerically. To determine this equation, we should initially solve the steady-state equations and find $\alpha_1$, $\alpha_2$, $\beta$, and $q_1$ to substitute in Eq. 5. It is noteworthy that the interactions of the OC mode and MW mode with MR generate some continuous variables (CV) entanglement such as the quantum correlation between cavities quadrature operators and MR position and momentum operator. For this reason, it is convenient to rewrite Eq. 5 based on OC and MW cavity fields quadrature fluctuations ($\delta X_i$, $\delta P_i$) and ($\delta X_b$, $\delta P_b$), respectively. Therefore, Eq. 5 in terms of fields' quadrature fluctuation becomes:

$$\dot{\delta q} = \omega_m \delta p, \quad \beta = \beta_r + i\beta_i, \alpha_1 = \alpha_{1r} + i\alpha_{1i}, \alpha_2 = \alpha_{2r} + i\alpha_{2i}, q_1 = q_{1r} + iq_{1i}$$

$$\dot{\delta p} = -\omega_m \delta q - \gamma_m \delta p + \sqrt{2}G_{o\omega}[\beta_r \delta X_b + \beta_i \delta P_b] + \sqrt{2}G_{oc}[\alpha_{1r}\delta X_1 + \alpha_{1i}\delta P_1] + \sqrt{2}G_{oc2}[\alpha_{2r}\delta X_2 + \alpha_{2i}\delta P_2] + \xi$$

$$\dot{\delta X_1} = -\kappa_c \delta X_1 + \Delta_c \delta P_1 - G_{oc}[q_{1r}\delta P_1 + q_{1i}\delta X_1] - \sqrt{2}G_{oc}\alpha_{1i}\delta q - 2\chi^{(2)}E_f[(\alpha_{2r}\delta P_1 - \alpha_{2i}\delta X_1) + (-\alpha_{1r}\delta P_2 + \alpha_{1i}\delta X_2)] + \sqrt{2\kappa_c}\delta X_{inc} \quad (7)$$

$$\dot{\delta P_1} = -\kappa_c \delta P_1 - \Delta_c \delta X_1 + G_{oc}[q_{1r}\delta X_1 - q_{1i}\delta P_1] + \sqrt{2}G_{oc}\alpha_{1r}\delta q - 2\chi^{(2)}E_f[(\alpha_{2r}\delta X_1 + \alpha_{2i}\delta P_1) + (\alpha_{1r}\delta X_2 + \alpha_{1i}\delta P_2)] + \sqrt{2\kappa_c}\delta P_{inc}$$

$$\dot{\delta X_2} = -\kappa_{c2}\delta X_2 + \Delta_{c2}\delta P_2 - G_{oc2}[q_{1r}\delta P_2 + q_{1i}\delta X_2] - \sqrt{2}G_{oc2}\alpha_{2i}\delta q - 2\chi^{(2)}E_f[-\alpha_{1r}\delta P_1 - \alpha_{1i}\delta X_1] + \sqrt{2\kappa_{c2}}\delta X_{inc2}$$

$$\dot{\delta P_2} = -\kappa_{c2}\delta P_2 - \Delta_{c2}\delta X_2 + G_{oc2}[q_{1r}\delta X_2 - q_{1i}\delta P_2] + \sqrt{2}G_{oc2}\alpha_{2r}\delta q - 2\chi^{(2)}E_f[\alpha_{1r}\delta X_1 - \alpha_{1i}\delta P_1] + \sqrt{2\kappa_{c2}}\delta P_{inc2}$$

$$\dot{\delta X_b} = -\kappa_\omega \delta X_b + \Delta_\omega \delta P_b - G_{o\omega}[q_{1r}\delta P_b + q_{1i}\delta X_b] - \sqrt{2}G_{o\omega}\beta_i\delta q + \sqrt{2\kappa_\omega}\delta X_{inb}$$

$$\dot{\delta P_b} = -\kappa_\omega \delta P_b - \Delta_\omega \delta X_b + G_{o\omega}[q_{1r}\delta X_b - q_{1i}\delta P_b] + \sqrt{2}G_{o\omega}\beta_r\delta q + \sqrt{2\kappa_\omega}\delta P_{inb}$$

where $\delta X_{inc} = (\varepsilon_c + \varepsilon_c^*)/\sqrt{2}$, $\delta P_{inc} = (\varepsilon_c - \varepsilon_c^*)/i\sqrt{2}$, $\delta X_{inc2} = (\varepsilon_{c2} + \varepsilon_{c2}^*)/\sqrt{2}$, $\delta P_{inc2} = (\varepsilon_{c2} - \varepsilon_{c2}^*)/i\sqrt{2}$, $\delta X_{inb} = (\varepsilon_b + \varepsilon_b^*)/\sqrt{2}$, $\delta P_{inb} = (\varepsilon_b - \varepsilon_b^*)/i\sqrt{2}$. Subsequently, Eq. 7 can be written in the compact form as:

$$\dot{u}(t) = A_{ij}u(t) + n(t), \quad (8)$$

The general form of Eq. 8 is re-written as follows:

$$\begin{bmatrix} \dot{\delta q} \\ \dot{\delta p} \\ \dot{\delta X_1} \\ \dot{\delta P_1} \\ \dot{\delta X_2} \\ \dot{\delta P_2} \\ \dot{\delta X_b} \\ \dot{\delta P_b} \end{bmatrix} = \underbrace{\begin{bmatrix} 0 & \omega_m & 0 & 0 & 0 & 0 & 0 & 0 \\ -\omega_m & -\gamma_m & G_{ocr} & G_{oci} & G_{ocr2} & G_{oci2} & G_{o\omega r} & G_{o\omega i} \\ -G_{oci} & 0 & A_1 & A_2 & -\chi_e^{(2)}\alpha_{1i} & \chi_e^{(2)}\alpha_{1r} & 0 & 0 \\ G_{ocr} & 0 & A_3 & A_4 & -\chi_e^{(2)}\alpha_{1r} & -\chi_e^{(2)}\alpha_{1i} & 0 & 0 \\ -G_{oci2} & 0 & \chi_e^{(2)}\alpha_{1i} & \chi_e^{(2)}\alpha_{1r} & A_5 & A_6 & 0 & 0 \\ G_{ocr2} & 0 & -\chi_e^{(2)}\alpha_{1r} & \chi_e^{(2)}\alpha_{1i} & A_7 & A_8 & 0 & 0 \\ -G_{o\omega i} & 0 & 0 & 0 & 0 & 0 & A_9 & A_{10} \\ G_{o\omega r} & 0 & 0 & 0 & 0 & 0 & A_{11} & A_{12} \end{bmatrix}}_{A_{i,j}} \times \underbrace{\begin{bmatrix} \delta q \\ \delta p \\ \delta X_1 \\ \delta P_1 \\ \delta X_2 \\ \delta P_2 \\ \delta X_b \\ \delta P_b \end{bmatrix}}_{u(t)} + \underbrace{\begin{bmatrix} 0 \\ \xi \\ \sqrt{2\kappa_c}\delta X_{inc} \\ \sqrt{2\kappa_c}\delta P_{inc} \\ \sqrt{2\kappa_{c2}}\delta X_{inc2} \\ \sqrt{2\kappa_{c2}}\delta P_{inc2} \\ \sqrt{2\kappa_\omega}\delta X_{inb} \\ \sqrt{2\kappa_\omega}\delta P_{inb} \end{bmatrix}}_{n(t)} \quad (9)$$

$G_{ocr} = \sqrt{2}G_{oc}\alpha_{1r}, G_{oci} = \sqrt{2}G_{oc}\alpha_{1i}, G_{ocr2} = \sqrt{2}G_{oc2}\alpha_{2r}, G_{oci2} = \sqrt{2}G_{oc2}\alpha_{2i}, G_{o\omega r} = \sqrt{2}G_{o\omega}\beta_r, G_{o\omega i} = \sqrt{2}G_{o\omega}\beta_i, \chi_e^{(2)} = 2\chi^{(2)}E_f$

$A_1 = -\kappa_c - G_{oc}q_{1i} + \chi_e^{(2)}\alpha_{2i}, A_2 = \Delta_c - G_{oc}q_{1r} - \chi_e^{(2)}\alpha_{2r}, A_4 = -\kappa_c - G_{oc}q_{1i} - \chi_e^{(2)}\alpha_{2i}, A_3 = -\Delta_c + G_{oc}q_{1r} - \chi_e^{(2)}\alpha_{2r}$

$A_5 = -\kappa_{c2} - G_{oc2}q_{1i}, A_6 = \Delta_{c2} - G_{oc2}q_{1r}, A_8 = -\kappa_{c2} - G_{oc2}q_{1i}, A_7 = -\Delta_{c2} + G_{oc2}q_{1r}$

$A_9 = -\kappa_\omega - G_{o\omega}q_{1i}, A_{10} = \Delta_\omega - G_{o\omega}q_{1r}, A_{12} = -\kappa_\omega - G_{o\omega}q_{1i}, A_{11} = -\Delta_\omega + G_{o\omega}q_{1r}$

As a simple form, Eq. 9 is solved as $u(t) = \exp(A_{i,j}t)u(0) + \int(\exp(A_{i,j}s).n(t-s))ds$, where $u(t)$ is a column matrix indicating the MR position and momentum operator and also optical and microwave cavities quadrature fluctuation operators, and $A_{ij}$ is an $8 \times 8$ matrix. Moreover, $n(s)$ is the noise column matrix. The contributed input noises obey the following correlation function [9], [11].

$$\langle \varepsilon_c(s)\varepsilon_c^*(s')\rangle = [N(\omega_c)+1]\delta(s-s'); \quad \langle \varepsilon_c^*(s)\varepsilon_c(s')\rangle = [N(\omega_c)]\delta(s-s')$$
$$\langle \varepsilon_{c2}(s)\varepsilon_{c2}^*(s')\rangle = [N(\omega_{c2})+1]\delta(s-s'); \quad \langle \varepsilon_{c2}^*(s)\varepsilon_{c2}(s')\rangle = [N(\omega_{c2})]\delta(s-s') \quad (10)$$
$$\langle \varepsilon_b(s)\varepsilon_b^*(s')\rangle = [N(\omega_\omega)+1]\delta(s-s'); \quad \langle \varepsilon_b^*(s)\varepsilon_b(s')\rangle = [N(\omega_\omega)]\delta(s-s')$$
$$\langle \xi(s)\xi^*(s')\rangle = 2\gamma_m[N(\omega_m)+1]\delta(s-s'); \quad \langle \xi^*(s)\xi(s')\rangle = 2\gamma_m[N(\omega_m)]\delta(s-s')$$

where $N(\omega) = [\exp(\hbar\omega/k_B T)-1]^{-1}$, and $k_B$ and $T$ stand for Boltzmann's constant and operation temperature, respectively [9, 11]. Indeed, $N(\omega)$ is the equilibrium mean thermal photon numbers of the different modes. By considering Eq. 10, one can clearly find that $N(\omega_{c2}) < N(\omega_c) < N(\omega_\omega) << N(\omega_m)$, which is due to the MR operation frequencies. Therefore, we can simply neglect the $N(\omega_{c2})$ and $N(\omega_c)$; however, $N(\omega_m)$ has a considerable amplitude and thus can be ignored in the low-temperature operation. An interesting aspect of such an opto-mechanical system is its stability. For this reason, one can check the $A_{ij}$ Eigenvalue to identify the system stability, having the knowledge that for a stable system all of the real parts of the contributed Eigenvalue should be negative [9-12, 19]. Since we tend to study the entanglement between modes in the developed system, we shall focus on the OC-MW modes, OC-MR modes, and MR-MW modes. It can be easily found in the various literatures that such bipartite entanglement can be quantified through Symplectic eigenvalue, which is given by [7], [8]:

$$\eta = \frac{1}{\sqrt{2}}\sqrt{\sigma \pm \sqrt{\sigma^2 - 2\det(\sigma)}}, \quad (11)$$
$$\sigma = \det(A) + \det(B) - 2\det(C)$$

where A, B, and C are 2×2 correlation matrix elements [A, C; $C^T$, D]. The entire correlation matrix elements can be easily constructed using $<\delta a_i \delta a_i>$, $<\delta a_i \delta a_i^+>$, $<\delta a_i^+ \delta a_i>$, $<\delta a_i^+ \delta a_i^+>$, $<\delta a_i>$, and $<\delta a_i^+>$, where "i" can be "m" for MR modes, "1" for OC modes, and "b" for MW modes. Finally, it should note that Eq.11 is an important criterion for identification of the entanglement between two modes. Using this equation, it has been found that if $2\eta > 1$ the considered modes are purely separable; otherwise, for $2\eta < 1$ two modes are entangled [8]. In the present study, we used this criterion to determine which of two modes are entangled, and which ones are not.

### III. SIMULATIONS AND RESULTS

In this section, we analyze the cavity modes entanglement with MR. All data presented in Table 1 were taken from Refs [9]-[12], [19]. Initially, to make a comparison with a traditional tripartite system, the second harmonic generation Hamiltonian and its coupling factor are considered zero; so, the developed system acts as a traditional opto-mechanical system. As shown in Fig. 2, by canceling all of the contributed factors of mode ($a_2$, $a_2^+$), the system acts like a traditional tripartite system (Sys I). In this figure, one can find the results for the entanglement between OC-MW, OC-MR, and MR-MW at T = 100 mK, and the effect of the MR mass. Interestingly, all of the simulated data in Fig, 2 are approximately comparable with results reported in earlier published works [9]-[11]. As can be seen from this figure, by increasing MR mass, the entanglement between modes is dramatically decreased; because by increasing the MR mass the coupling factors such as $G_{oc}$ and $G_{o\omega}$ are decreased. Additionally and importantly, the effect of the temperature changing is considered for the Sys I in terms of m = 20 ng (Fig. 3). It is identified that by increasing the temperature, all of the cavity modes entanglements are severely decreased. Such a drop is contributed to the noise factor, which is straightforwardly dependent on environment temperature and frequency. From Eq. 10, we notice that $N(\omega_m)$ has a considerable value when the temperature increases; thus, the noise is introduced due to the MR system influence on the coupled OC mode and MW cavity mode. The results illustrated in Fig. 3 show that also the entanglement between MW-OC and MW-MR is affected by the thermal photon number of the introduced fields. In line with the published works [9]-[12], [19], we found that entanglement in OC-MW is easily affected by the MR introduced noise, so we have to engineer and improve the opto-mechanical tripartite system to solve the problem. Moreover, for a better understanding of some critical points around $2\eta = 1$ in Fig. 3a and 3c, we inserted its magnified picture. In these inset figures, one can clearly trace the entanglement changing. In the following, we focus on the OC-MW cavities mode entanglement, because of its importance in the illumination system used in the quantum radar sensing system. Actually, entanglements between MW cavity modes and OC mode are mostly used in quantum radar [1]. In this section, we compared Sys I and Sys II (Fig. 4). As presented in Fig. 4, by considering a similar condition for the traditional tripartite system and developed one, we found that by introducing modes $a_2$ in the system the OC-MW entanglement profile is slightly changed. However, the system behavior in the low temperature is similar to each other. In fact, it is observed that at three different frequencies such as $\Delta\omega/\omega \sim -1, 0$, and 1 the entanglement between cavity modes is the maximum. Modeling results have proved that in the low-temperature condition the introduced systems (Sys I and Sys II) are operated approximately in a similar way. However, our article originally is attributed to the fact that the developed system can perfectly lead to subside the MR noise-introducing. In the following, we try to found how Sys II can solve this problem without changing the MR bandwidth engineering.

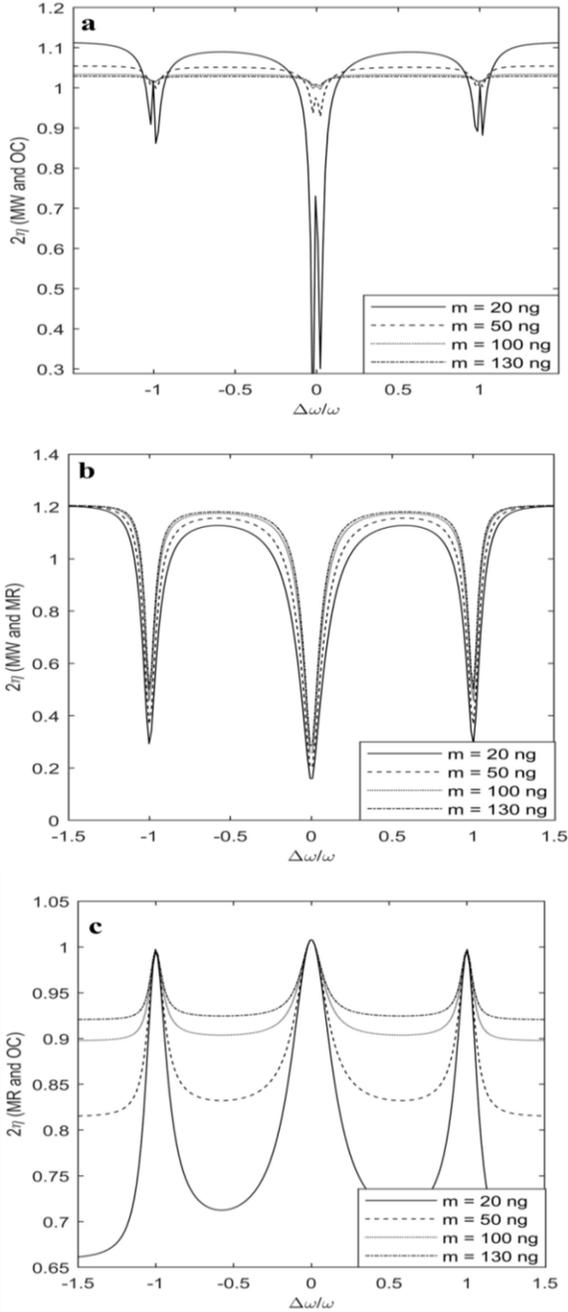

Fig. 2, MR mass effect on cavity modes entanglement vs $\Delta\omega/\omega$;
(a) MW-OC, (b) MW-MR, (c) MR-OC, T = 100 mK.

Fig. 5 presents the operation of the developed system for the case of environment temperature rise. The entanglement of the OC-MW and MW-MR is considered in Figs. 5a and 5b. Their magnified critical areas and the related inset figures show the related cases. A remarkable result depicted in Fig. 5a is that where the temperature increases up to 10 K, it does not affect the OC-MW entangling, while when the temperature reaches 100 K, OC-MW becomes pure and separable. The concluding remark of this work is that using the developed opto-mechanical system, the OC-MW entangling is guaranteed for the temperature smaller than 50 K. Moreover, a similar study was done for MW-MR modes entanglement and found that for the environment temperature smaller than 5 K, the MW-MR modes remained entangled. It has to be noted that the separability increase rate in Sys II is so smaller than Sys I. To prove this point, one can compare the Figs. 3c and 5b. This result can be attributed to the spherical capacitor embedded in OC that is parallel to MW circuit capacitor. Indeed, by MW circuit changing, we did an important task by which OC mode can be coupled to MW cavity mode directly, leading to an interesting key to decline the MR noise action on system performance.

Table I: Constants used in this study [9-12]

| | |
|---|---|
| $\lambda_c$ (incidence wavelength) | 808 nm |
| $f_m$ (MR resonance frequency) | 10 MHz |
| $f_\omega$ (MW resonance frequency) | 10 GHz |
| $L_c$ (OC length) | 1 mm |
| $P_c = P_\omega$ (Cavities driving power) | 30 mW |
| $\Delta_c = \Delta_{c2}$ (OC detuning factors) | $\omega_m = 2\pi f_m$ |
| $\gamma_m$ (MR damping rate) | 110 |
| $\kappa_c$ (OC damping rate) | $0.02\omega_m$ |
| $\kappa_{c2}$ (Sphere plates damping rate) | 1e13 (1/s) |
| $\kappa_\omega$ (MW damping rate) | $0.03\omega_m$ |
| m (MR resonator mass) | 20 ng |
| T (environment temperature) | 100 mK |

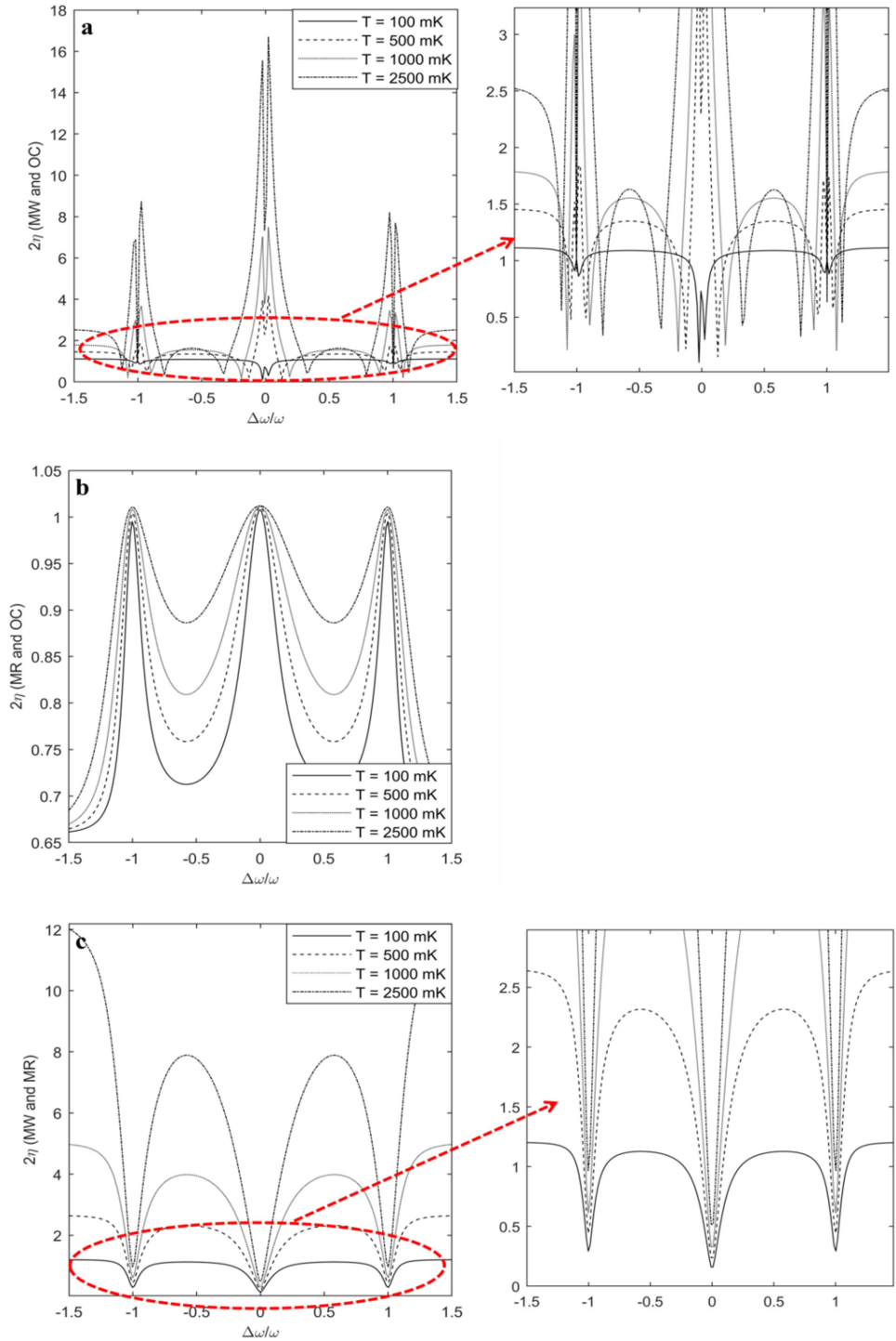

Fig. 3 Temperature effect on cavity modes entanglement vs $\Delta\omega/\omega$; (a) MW-OC, (b) OC-MR, (c) MR-MW, m = 20 ng.

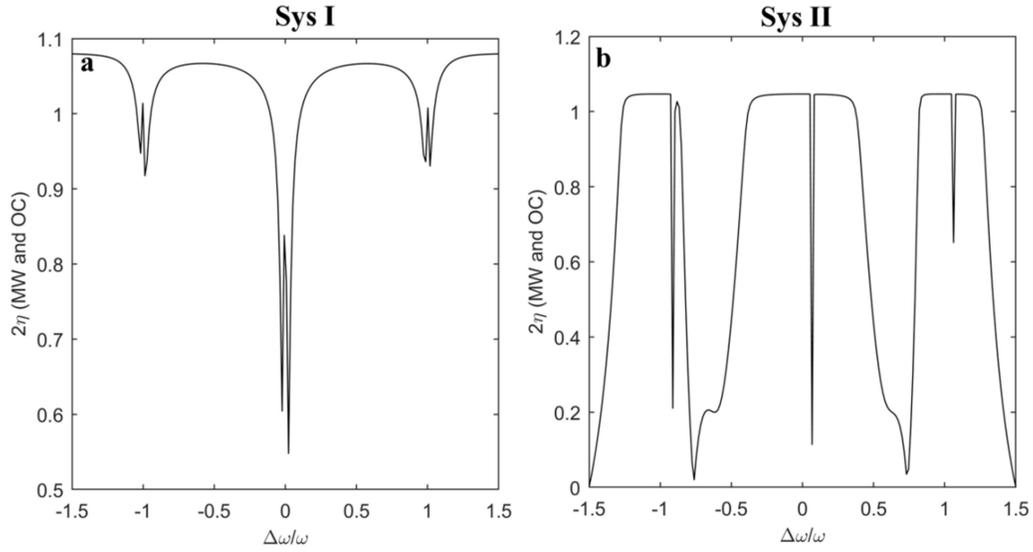

Fig. 4 Comparison between entanglement of MW-OC for: (a) Sys I, (b) Sys II at T = 100 mK, m = 20 ng.

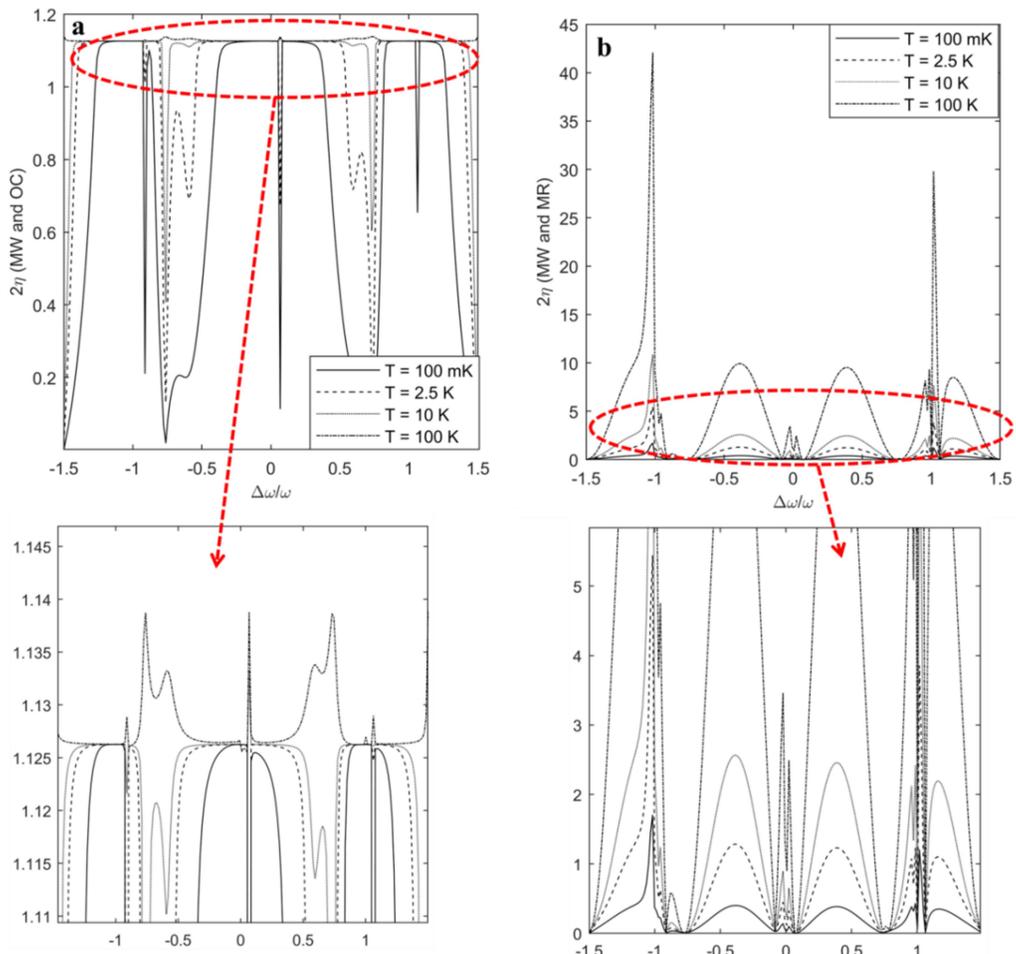

Fig 5. Temperature effect on the cavity modes entanglement in Sys II, (a) MW-OC, (b) MW-MR, m = 20 ng.

## IV. CONCLUSIONS

In this work, we introduced a developed opto-mechanical system in which a spherical capacitor is embedded into the optical cavity. By this alteration, a new mode is generated through the plasmon-plasmon interaction in the capacitor gap region. The improved tripartite system was analyzed quantum mechanically and the related dynamics of motion were derived using Heisenberg-Langevin equations. In the following, the entanglement between optical and microwave cavity modes and microresonator mode were investigated. It was revealed that by changing the system specifications such as microresonator mass, the modes entanglement were changed, probably due to the system coupling factors alteration. Furthermore, as the concluding remark of the present work, we showed that by changing the temperature, the entanglements between modes are dramatically changed. Thus, by the modeling results, it is shown that the developed system is able to retain modes entanglement when the environment temperature reaches around 50 K. We know from the literature that by a traditional opto-mechanical system we did not reach such a temperature. However, it should again be noted that some published works used microresonator bandwidth engineering to solve the temperature problem.